\documentclass[twocolumn,showpacs,aps,prl,superscriptaddress]{revtex4}
\usepackage[paperwidth=210mm,paperheight=297mm,centering,hmargin=2cm,vmargin=2.5cm]{geometry}
\usepackage{amssymb}
\usepackage{graphicx}
\usepackage{dcolumn}
\usepackage{amsmath}
\usepackage{latexsym}
\usepackage{amsbsy}
\usepackage{bm}
\usepackage{subfigure}
\begin {document}

\title {Strength of Fractured Rocks}
\author {Chandreyee Roy}
\email{chandreyee.roy@bose.res.in}
\noaffiliation
\affiliation{
\begin {tabular}{c}
Satyendra Nath Bose National Centre for Basic Sciences, 
Block-JD, Sector-III, Salt Lake, Kolkata-700098, India
\end{tabular}
}
\author{Srutarshi Pradhan}
\email{srutarshi@sintef.no}
\noaffiliation
\affiliation{
SINTEF Petroleum Research, S.P. Anderseng veg 15 B, 7031, Trondheim, Norway
}
\author{Anna Stroisz}
\email{Anna.stroisz@sintef.no}
\noaffiliation
\affiliation{
SINTEF Petroleum Research, S.P. Anderseng veg 15 B, 7031, Trondheim, Norway
}
\author{Erling Fj\ae r}
\email{Erling.Fjaer@sintef.no}
\noaffiliation
\affiliation{
SINTEF Petroleum Research, S.P. Anderseng veg 15 B, 7031, Trondheim, Norway
}

\begin {abstract}
In this report we present a study on the strength of rocks which are partially fractured from before. We have considered a two dimensional case of a rock in the form of a lattice structure. The fiber bundle model is used for modelling the $2-D$ rock. Each lattice site is considered to be a fiber which has a breaking threshold. Fractures in this system will be of the form a cluster of sites and the length is defined as the number of sites belonging to a single cluster. We introduce fractures in the system initially and apply load until the rock breaks. The breaking of a rock is characterized by a horizontal fracture which connects the left side of the lattice to the right side. The length distribution and the strength of such systems have been measured.  
\end {abstract}
\maketitle

\section{1. Introduction}
\begin{figure}
\includegraphics[scale=0.3]{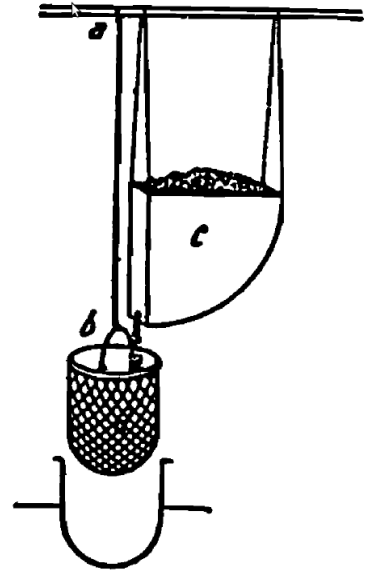}
\caption{ Tensile strength experiment by Leonardo Da Vinci \citep{davinci}}
\label{davinci} 
\end{figure}

This report is on the study of strengths of fractured rocks. Some breakdown properties of such rocks have also been studied. For the theoretical analysis, the simple Fiber Bundle Model and the Discrete Fracture Network have been considered. A Fiber Bundle Model (FBM) is used in material sciences to study the breakdown properties of materials. Studying such properties of materials had been first introduced by Leonardo Da Vinci about five hundred years ago. In one of his notebooks he describes an experiment to measure the strength of wires as a function of their lengths. He attached a bucket at one end of a wire and clamped the other end (see Fig. \ref{davinci}). Sand was allowed to pour into the bucket until the wire broke. A small pit was created just below the bucket so that when the wire broke, it fell into the pit. The weight of the sand inside the bucket was used to measure the tensile strength of the wire. He found that the longer the wires are, the weaker they are. 
\begin{figure}
\centering
\includegraphics[scale=0.30]{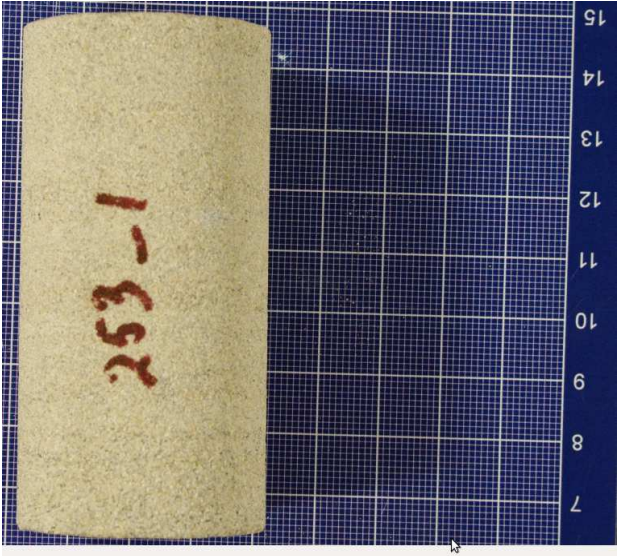}
\caption{Lateral view of the Castle Gate Rock sample.}
\label{cgrock}
\end{figure}

Generally, one has always been interested in material stability. For example, when people build houses they try to make it in such a way so that it can withstand normal weather conditions.  The Fiber Bundle Model was first introduced by Pierce \cite{pierce} in 1926 to test the strength of cotton yarns. Since then many people have worked on various aspects of it resulting in such a simple model to have a vast literature today. These models are perfect for studying their failure phenomena as a part of theoretical physics. Several successful experiments also have been carried out. The Discrete Fracture Network (DFN) model is used extensively in studying fractures. It was first introduced by Darcel et al \citep{darcel-davy} in 2003. It essentially captures the properties of a real fracture network.

\begin{figure}
\includegraphics[scale=0.300]{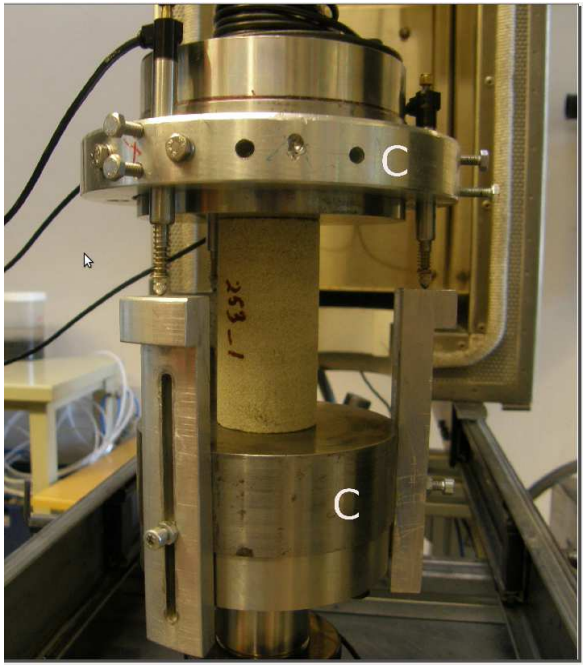}
\caption{Experimental setup}
\label{exptsetup}
\end{figure}

\begin{figure}[b]
\centering
\includegraphics[scale=0.25]{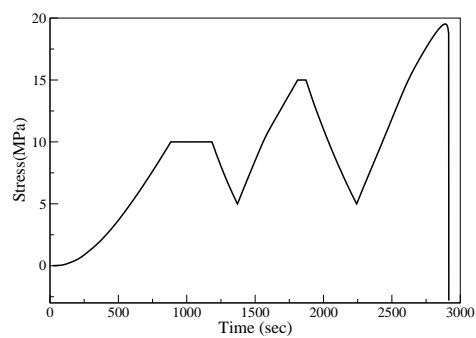}
\caption{ Stress - Time curve}
\label{stresstime}
\end{figure}

\section{2. Experimental Data}

\begin{figure}
\centering
\includegraphics[scale=0.3]{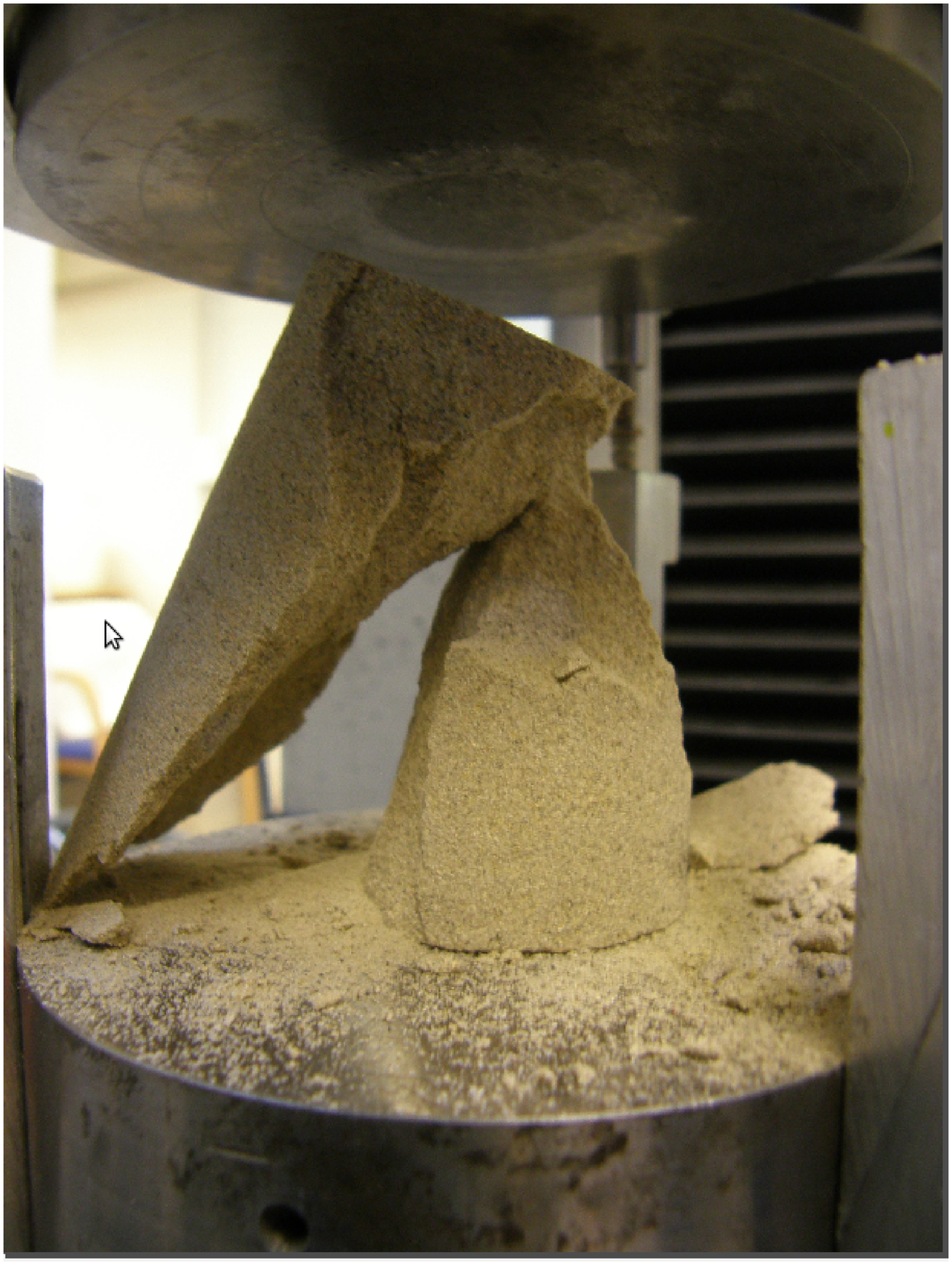}
\caption{ Sample after failure}
\label{failure}
\end{figure}

\begin{figure}[b]
\centering
\subfigure[]{
\includegraphics[scale=0.25]{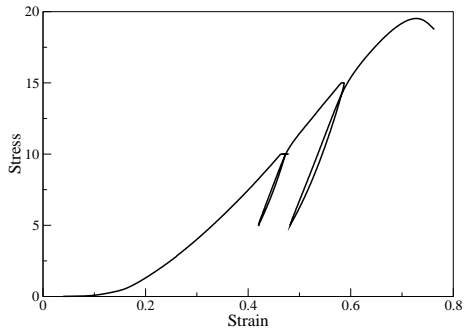}
} \\
\vspace{0.3cm}
\subfigure[]{
\includegraphics[scale=0.25]{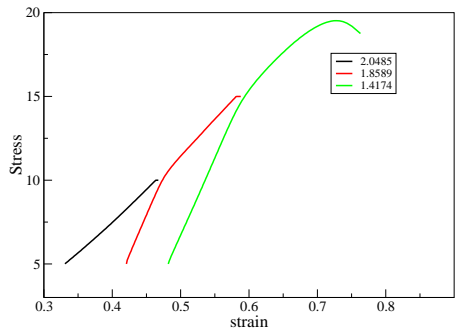}
}
\caption{ Stress - Strain Curve}
\label{stressstrain}
\end{figure}

The modified unconfined compressive strength (UCS) test was conducted on the Castlegate sandstone. 
The sample was cored perpendicular to bedding and cut to size of approximately 77.62 mm in diameter
and 38.2 mm in length. The lateral view of core is shown in Fig. \ref{cgrock}. The experiment was performed using a servo-controlled loading frame (Fig.\ref{exptsetup}). The sample was placed between two pistons, a movable upper piston and immovable base, marked with $C$ on the figure. Stress was applied in the vertical direction only, and the axial deformation was measured with three LVDTs fixed around the sample. The failure, unlike the standard UCS test, was achieved by a stepwise loading with $5$ MPa intervals. Every new stress level was preceded by unloading to $5$ MPa (see Fig. \ref{stresstime} for the stress path). The procedure was repeated until the rock failure (Fig. \ref{failure}). The stress versus strain curve for this experiment is shown in Fig. \ref{stressstrain}(a). This figure has been re-plotted in Fig. \ref{stressstrain}(b) after eliminating the unloading process of the experiment.  


\section{2. Fiber Bundle Models}
\begin{figure}
\centering
\subfigure[]{
\includegraphics[scale=0.3]{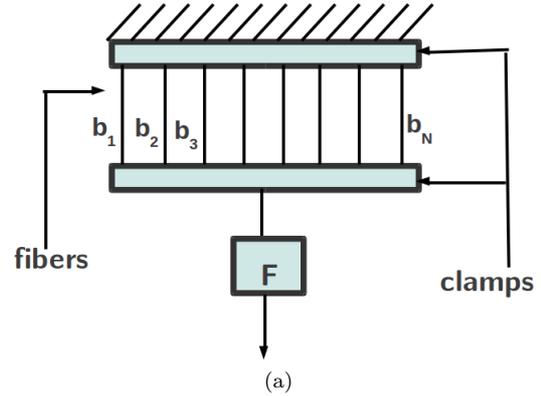}
}
\subfigure[]{
\includegraphics[scale=0.3]{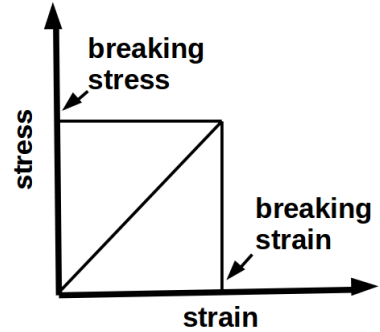}
}
\caption{(a)Schematic diagram of the Fiber Bundle model, (b)Stress strain relation (Hooke's Law)}
\label{fbm}
\end{figure}

A Fiber Bundle Model is basically a set of elastic fibers which are placed parallely one after the other as shown in Fig. \ref{fbm}(a). The fibers are clamped at both ends. The top end is fixed to a rigid support and at the lower end an external force is applied to elongate the bundle. This external force is distributed equally among all intact fibers. Each fiber has a unique breaking threshold ${b_i}$. All the fibers follow Hooke's Law until the load acting on each fiber reaches their respective breaking thresholds (see Fig. \ref{fbm}(b)). The elastic constant is assumed to be unity. So the stress applied to each fiber is equal to the elongation caused in it. Each fiber is also assumed to be brittle which means that if the external load per fiber is equal to the threshold value of the fiber then it immediately breaks off. 

On the application of an external load to a fiber bundle, the fibers having threshold values lower than the acting load per fiber break. When a fiber breaks, it releases the stress carried by it. This is described as stress relaxation. The released stress will now be distributed among the remaining intact fibers. There exists many ways in which the released stress can be shared. Depending on the redistribution of released stress various models of the Fiber Bundle exist in the literature. If the released stress is distributed to all the remaining intact fibers, then such a model is called Equal Load Sharing Model (ELS). On the other hand, if the released stress is distributed to only the neighbouring intact fibers of the broken fiber, then the model is called Local Load Sharing Model (LLS). The LLS model is more realistic than ELS. However, to model the strength of fractured rocks, we have considered the simple ELS rule of load sharing. 

When the fibers break on the application of an external load, the stress acting on the remaining intact fibers get enhanced. This new enhanced stress may now become more than the threshold values of some of the intact fibers. This results into the breakage of more fibers. This process continues till a stable state is reached. A stable state is described as a state where the threshold values of all the fibers is less than the stress per fiber acting at that particular time. It can also be described as a state when all the fibers have broken which occurs only if the external load per fiber is greater than the critical stress. A review of the model can be found in \citep{pradhan-hansen}.

\section{3. Modelling the fractures}
We have considered a two dimensional case of a fractured rock. We have used the following models for this.  
\subsection{3.1. Equal Load Sharing Model (ELS)}
The Equal Load Sharing Model is also known as the Global Load sharing (GLS) Model. It is the simplest and the oldest of all the Fiber Bundle Models. When external load is applied to the bundle then all the fibers having their breaking thresholds less than the external load per fiber will break. In this model the stress released by these broken fibers will be distributed globally and equally among all the remaining intact fibers. This means that if a fiber breaks then the load released by it may affect any other fiber which can be at an infinite distance from the breaking fiber. Thus each fiber has an infinite range of interaction. ELS is a mean field type model and it is easier to attack this model analytically than the other models. Around 60 years ago Daniel, 1945 \cite{daniels} had given some exact analytical results of this model. We have used this model for our analysis of the strength of fractured rocks.

\subsection{3.2. Discrete Fracture Network (DFN)}
    A Discrete Fracture Network (DFN) model was introduced by Darcel {\it et al} \citep{darcel-bour,darcel-davy,darcel-odling} in the year 2003. The model captures the essential features of fracture outcrops which occur naturally in nature. It can be constructed in two and three dimensions. The model basically incorporates the properties that the length of fractures that happen in nature broadly follow power law and the position of fracture centres are heavily fractal in nature. 

\subsection{3.3. Our Model}
   We have considered two dimensional square lattice of size $L \times L$ where $L$ is the number of sites in each column or row. We have considered only one property of the Discrete Fracture Network Model for now which is the power law distributed lengths of the fractures. (In a future work we plan to include the fractal nature of the fracture centres). Each lattice site is considered to be a fiber having a particular strength. There is no periodic boundary condition in the horizontal and vertical directions. Each fiber (i.e. each site) is assigned a threshold value of strength that it can endure. These values are taken from a uniform distribution between $[0:1]$.
   
\begin{figure}
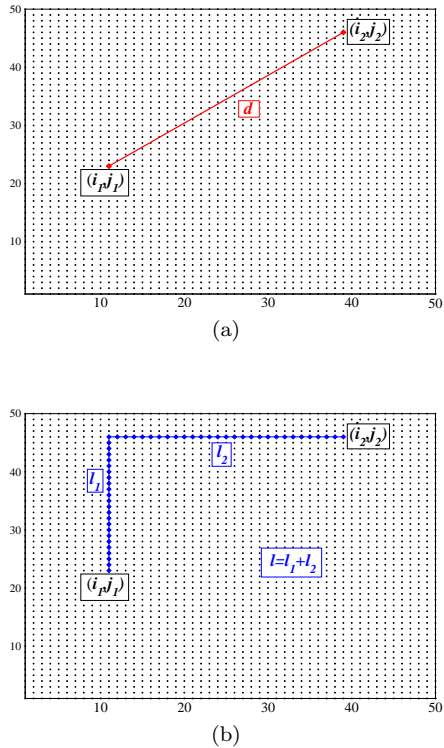

\centering
\subfigure[]{
\includegraphics[scale=0.25]{figure8a.eps} 
} \\
\vspace{0.5cm}
\subfigure[]{
\includegraphics[scale=0.25]{figure8b.eps}
}
\caption{ Process of creating the fractures on a square lattice}
\label{fracprocess}
\end{figure}

   The fractures are dropped in the lattice in the following manner. Two points are chosen randomly on the lattice and a straight line is drawn which connects these two points (see Fig. \ref{fracprocess}(a)). The length of the straight line $d$ is calculated. Since the fracture lengths follow a power law the probability of finding a fracture of length $d$ is proportional to $d^{-a}$ where $d$ is the length of a fracture and $a$ is the slope of the power law. This implies that 
  \begin{equation}
    P(d) = Const . d^{-a} 
  \end{equation}

\begin{figure}[t]
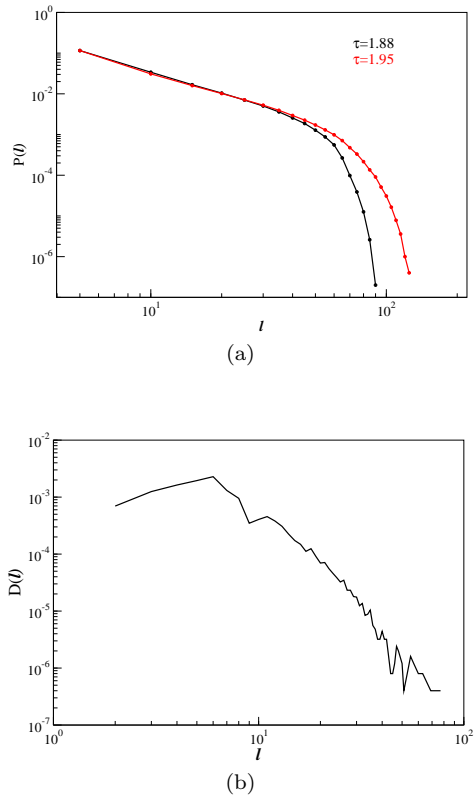

\centering
\subfigure[]{
\includegraphics[scale=0.25]{figure9a.eps}
}\\
\vspace{0.5cm}
\subfigure[]{
\includegraphics[scale=0.25]{figure9b.eps}
}
\caption{(a) Length distribution of the fractures for (i) length given by $d$ (black) (ii) length given by $l$ (red). Lattice length $L=64$, number of fractures $=100$, number of configurations $=10000$. (b) Length distribution after taking into account the length of merged fractures. Upper bound of length $=5 units$, number of fractures $=50$, number of configurations $=10000$. }
\label{lengthdistributuion}
\end{figure}

 A random number $r$ is chosen from a uniform distribution between $[0:1]$. If $r$ is less than or equal to $P(d)$, then we keep the straight line or else we discard it and choose another set of two points. This ensures that the lengths of the straight lines follow a power law. To place the fractures on the lattice, we connect the chosen set of two points by passing through the sites in either a horizontal or vertical manner. This process is depicted in Fig. \ref{fracprocess}(b). The length of a fracture in this case is defined by the number of sites included in a particular fracture while moving from point $(i_1,j_1)$ to point $(i_2,j_2)$ given by $l=l_1+l_2$. The length distribution of these lattice fractures for both the lengths $d$ and $l$ are plotted in Fig. \ref{lengthdistributuion}(a). From the figure one can see that there is not much difference between the two power laws.
 
 However, it may happen that while placing the fractures, two fractures merge together to give one fracture. In that case the power law changes as shown in Fig. \ref{lengthdistributuion}(b). Here, an upper bound in the length of fractures has been maintained such that no long fractures are formed connecting one side of the lattice to the other.

\begin{figure}[b]
\includegraphics[scale=0.25]{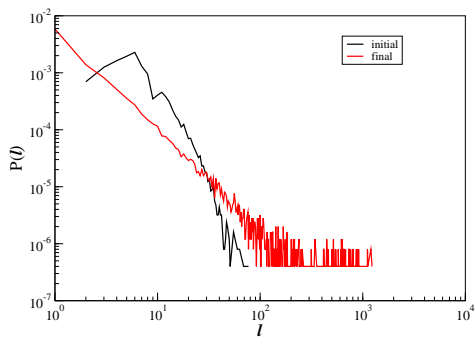}
\caption{ Length distribution of fractures before loading (black) and after the $2-D$ rock has broken (red)}
\label{lengthdistributioninitialandfinal}
\end{figure}
\begin{figure}[b]
\centering
\subfigure[Upper bound for fracture lengths = 5 units]{
\includegraphics[scale=0.25]{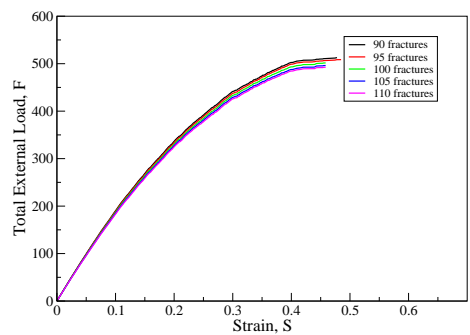}
}\\
\vspace{0.4cm}
\subfigure[Upper bound for fracture lengths = 10 units]{
\includegraphics[scale=0.25]{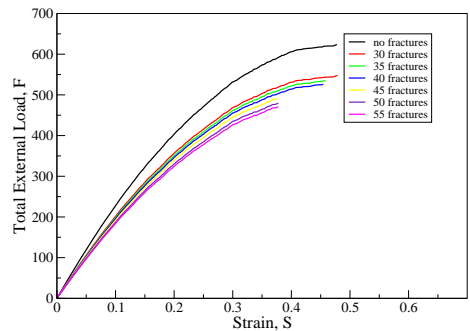}
}
\caption{ Avalanche dynamics: Total external load vs strain for different bounds of the fracture lengths.}
\label{avalanchedynamics}
\end{figure}

\section{4. Model Dynamics and Length Distribution}
The definition of an avalanche as described by Hemmer {\it et al} \cite{hemmer-hansen,hansen-hemmer,kloster-hansen,Sornette} is as follows. Let the number of sites or fibers broken due to an avalanche be given by $k$. The remaining number of intact sites are $L^2-k$. We first manually break the weakest site from the intact sites. The load released by this weakest site is then redistributed among all the other intact sites. If no other site breaks due to the enhanced stress per fiber, then it is called an avalanche of size $1$. On the other hand if more fibers break, then this triggers an avalanche of broken fibers of sizes more than $1$. After the system has reached a stable point we then find out the weakest site among the remaining intact sites and increase the external stress just enough such that only the weakest one breaks. Due to this another avalanche may or may not follow. For each increase in load we find out the avalanche size which is the number of sites that break (or fail) caused by the increased load. We carry out this process until we can find a horizontal path of broken sites which divides the system into a minimum of two distinct parts. We define the external load at this stage to be the strength of the rock.Fig. \ref{lengthdistributioninitialandfinal} shows the length distribution of the fractures before starting the avalanche (black) and after the last avalanche (red). The slope of initial length distribution is $3.062$ and for the final length distribution, it is $1.778$. Fig. \ref{avalanchedynamics} shows the total external load versus the strain that is applied to the bundle. In Fig. \ref{avalanchedynamics}(a) the upper bound is $5$. Thus a higher number of fractures have to be placed initially to get horizontal fracture connecting the left side of the lattice to the right side. In Fig. \ref{avalanchedynamics}(b) the upper bound is $10$ and so the number of initial fractures required is less.

\begin{figure}
\includegraphics[scale=0.25]{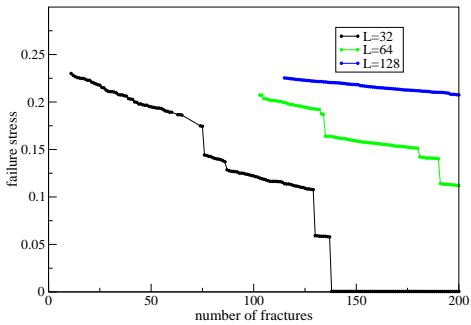}
\caption{ Stress Vs number of initial fractures for lattice size $L=32,64,128$.}
\label{stressfrac}
\end{figure}

\begin{figure}[b]
\centering
\subfigure[]{
\includegraphics[scale=0.25]{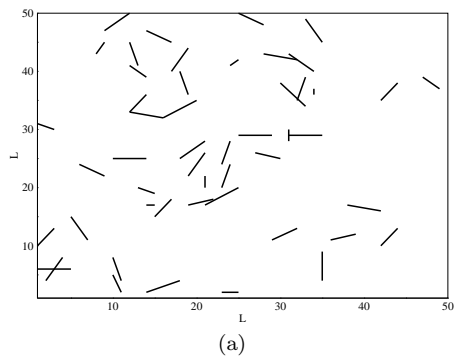}
}\\
\vspace{0.4cm}
\subfigure[]{
\includegraphics[scale=0.25]{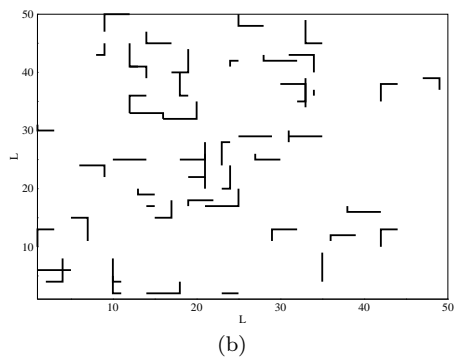}
}
\caption{ Snapshot of Initial fractures (a) without including lattice sites and (b) including lattice sites. Here length of lattice $L=50$, number of fractures placed $=50$, upper bound of fracture lengths $=5$.}
\label{initialfracpic}
\end{figure}

\begin{figure}
\centering
\subfigure[Fracture at percolation point (red)]{
\includegraphics[scale=0.25]{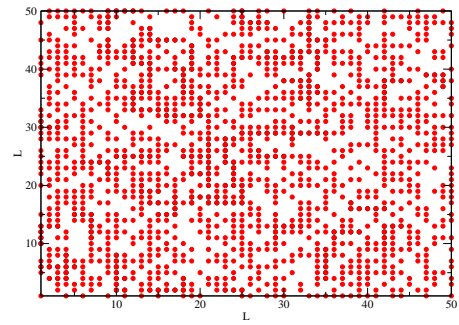}
}\\
\vspace{0.4cm}
\subfigure[Largest cluster at percolation point (green)]{
\includegraphics[scale=0.25]{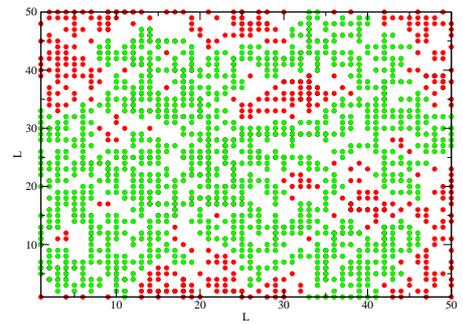}
}\\
\vspace{0.4cm}
\subfigure[Largest cluster just before (purple) and at (blue) percolation point]{
\includegraphics[scale=0.25]{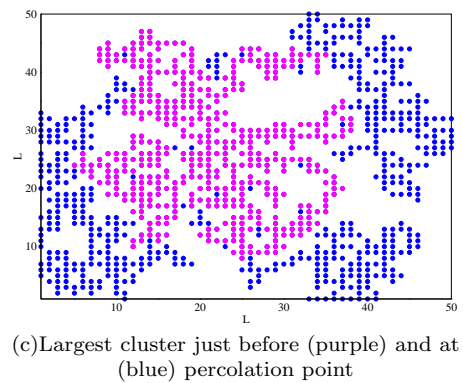}
}
\caption{ Snapshot of fractures starting from the initial loading upto the breaking point (percolation point)}
\label{percpointpic}
\end{figure}

\begin{figure}
\includegraphics[scale=0.25]{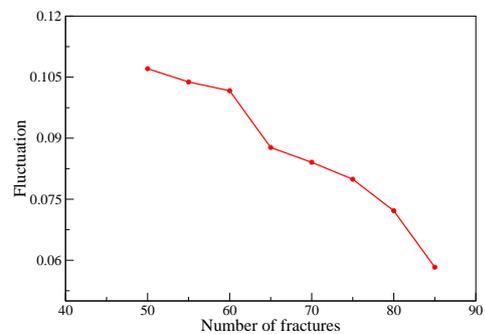}
\caption{ Fluctuation or Standard deviation with respect to number of initial fractures }
\label{fluctuation}
\end{figure}
\section{5. Strength of fractured rocks}
The strength of a fractured rock is defined as the external load at which a fracture is created which connects the left side of the lattice to the right side. The stress per fiber is calculated and plotted with respect to the number of initial fractures in Fig. \ref{stressfrac} for different $L$ values. The sudden drops in the plot for $L=32$ and $L=64$ indicate that when the number of initial fractures are sufficiently increased then there already exists a large cluster of broken fibers which on further loading percolates from the left of the lattice to the right of the lattice. When the strength is of the order of $0$, it means that the initial number of fractures is so high that the rock is already broken. For $L=128$ the sudden drop is expected to appear if the fracture numbers are increased even more than $200$. Fig. \ref{initialfracpic}(a) shows an initial configuration of the placement of initial fractures without including the lattice sites. The upper bound in the length of the fracture was given to be $5$ units. Fig. \ref{initialfracpic}(b) shows the initial configuration of fractures considering the lattice sites. Fig. \ref{percpointpic}(a) shows the state of the lattice at the breaking point. Red dots indicate broken lattice sites. Green dots in Fig. \ref{percpointpic}(b) represent the largest cluster at breakdown. In Fig. \ref{percpointpic}(c) we have shown how the largest cluster evolved. Purple dots represent the largest cluster just before breakdown and (purple $+$ blue) dots represent largest cluster at breakdown. Fig. \ref{fluctuation} shows the fluctuation of the strengths of the fractured rocks when the number of initial fractures are increased. 

\section{6. Conclusion}
To model a fractured rock in $2-D$, a $2-D$ lattice of length $L$ is considered where each site is assumed to be a fiber. Each site has its own breaking threshold value drawn from a uniform distribution between $[0:1]$. Cracks are applied in the form of broken sites and equal load distribution is carried out until the sample breaks. A sample is assumed to be broken when a fracture is created which connects the left side of the lattice to the right side. The strength of the rock is defined to be the load per fiber at which the sample breaks. The strength was plotted with respect to the number of initial fractures. We notice sudden drops in the strength which is due to the large clusters of sites being formed in the lattice. These large clusters weaken the lattice. The length distributions of the fractures were also plotted.

\begin{figure}
\centering
\subfigure[]{
\includegraphics[scale=0.25,angle=270]{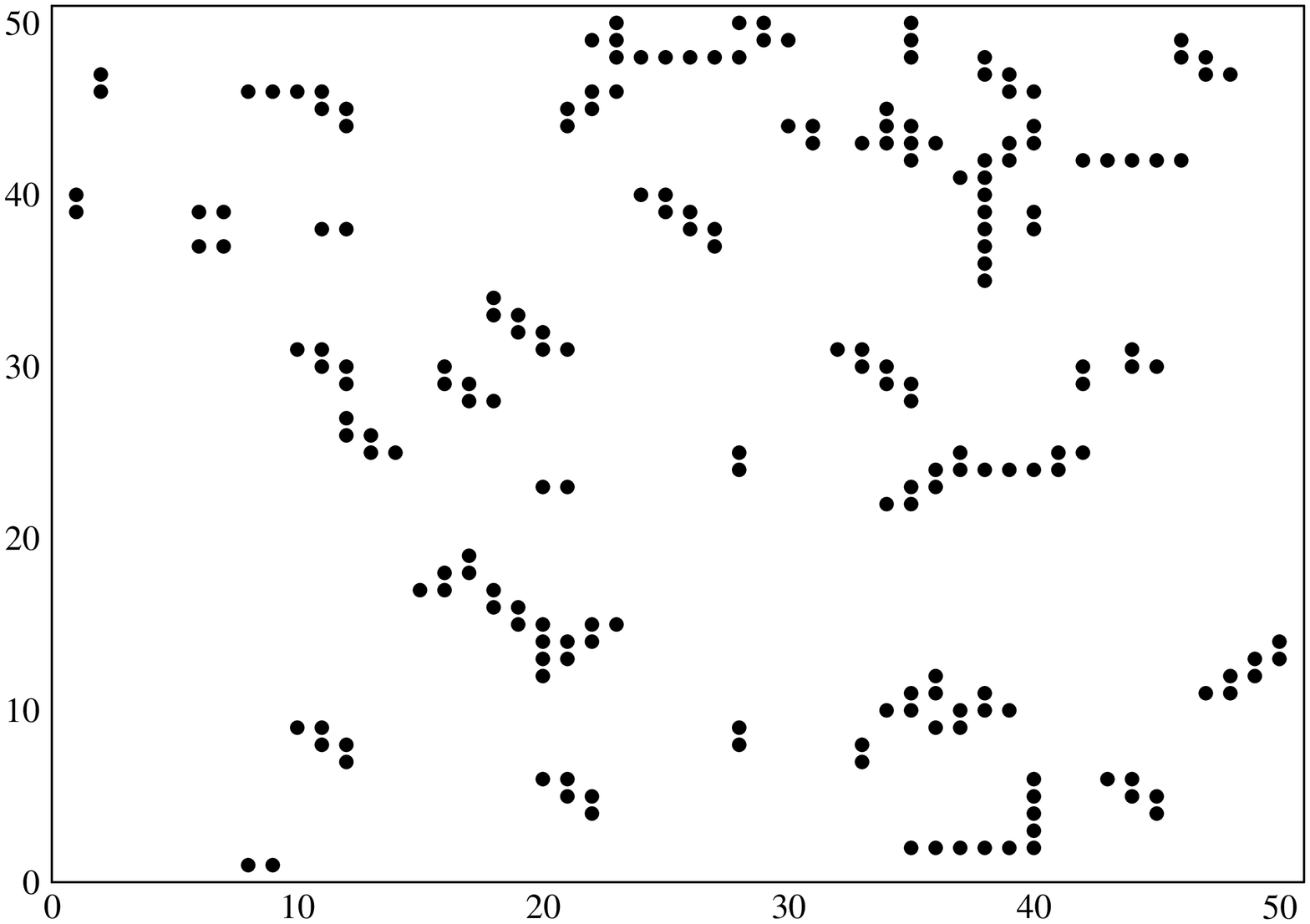}
}\\
\vspace{0.4cm}
\subfigure[]{
\includegraphics[scale=0.25]{figure16b.eps}
}
\caption{ (a) Snapshot of the initial fractures placed on the lattice before loading (b) Snapshot of the broken lattice. Magenta filled circles indicate broken lattice points and black filled circles indicate the initial position of the fractures for a lattice of size $L=50$.}
\label{nobreak}
\end{figure}

\section{7. Thoughts for Future Plans}
 The model presented in this report is a very simplified one. Some modifications are needed in this model to make it more realistic. Initially we were breaking the diagonal fractures into horizontal and perpendicular lines as shown in Fig. \ref{fracprocess}. Now, instead of carrying out the aforesaid process the diagonal fractures can be constructed in such a manner such that they fall on the lattice sites and remain diagonal as well as shown in Fig. \ref{nobreak}(a). Fig. \ref{nobreak}(b) shows the final state of the broken lattice when load is applied only around the fracture ends. We will be carrying out the same analysis for such a system as done here in the report. Then we will take into account the fractal dimension of the fracture centres while creating the DFN model and carry out the same analysis. We are also planning to include local load sharing dynamics into the model such that the load distribution is more localized. In this case the sites which are neighbours of the broken sites will have a greater probability to break which is more realistic in nature. 

\section{Acknowledgement}
The research work in this paper is a result of the scientific collaboration in
the INDNOR project 217413/E20 funded by the Research Council of Norway.

\end {document}